\documentclass[10pt,conference,letterpaper]{IEEEtran}
\IEEEoverridecommandlockouts

\usepackage[T1]{fontenc}
\usepackage{cite}
\usepackage{amsmath,amssymb,amsfonts}
\usepackage{graphicx}
\usepackage{textcomp}
\usepackage{xcolor}
\usepackage{booktabs}
\usepackage{array}
\usepackage{url}
\graphicspath{{figures/}}

\usepackage[colorlinks=true,urlcolor=blue,linkcolor=black,citecolor=black,
            pdfauthor={},pdftitle={Adaptive Warm-State Retention for Serverless Cold-Start Mitigation}]{hyperref}
\def\BibTeX{{\rm B\kern-.05em{\sc i\kern-.025em b}\kern-.08em
    T\kern-.1667em\lower.7ex\hbox{E}\kern-.125emX}}

\title{Adaptive Warm-State Retention for Serverless\\Cold-Start Mitigation: Treating Keep-Alive Time\\as an Online Control Parameter}

\author{%
  \IEEEauthorblockN{Sanjeet Raj Pandey}
  \IEEEauthorblockA{Service-centric Networking\\
    Technische Universit\"at Berlin\\
    Berlin, Germany\\
    s.pandey@tu-berlin.de}
  \and
  \IEEEauthorblockN{Axel K\"upper}
  \IEEEauthorblockA{Service-centric Networking\\
    Technische Universit\"at Berlin\\
    Berlin, Germany\\
    axel.kuepper@tu-berlin.de}
}

\begin{document}
\maketitle

\begin{abstract}
Serverless platforms are easy to operate but pay a recurring price in
\emph{cold starts}. When a function is called after sitting idle, its execution
environment has to be rebuilt before the request can run, and response time
suffers. The usual fix is to keep idle environments alive for a fixed window,
yet a single fixed value rarely fits a changing workload. Set it too high and idle
capacity is wasted; set it too low and avoidable cold starts increase. We argue that
the keep-alive window should not be a fixed deployment constant. We instead treat it
as a per-function runtime parameter that is adjusted during execution from what the
system actually observes, and we keep this question separate from how many replicas
to run.
To test the idea we built an external retention controller for OpenFaaS/faasd
that needs no changes to the platform, and drove it with two deliberately simple
agents: a reactive Additive-Increase Multiplicative-Decrease (AIMD) rule and an
Upper-Confidence-Bound (UCB1) multi-armed bandit. We evaluated them against
fixed-timeout baselines and a predictive histogram keep-alive on steady, bursty, and
trace-derived workloads (Huawei Serverless 2023), measuring cold-start rate, latency,
and accumulated warm time. The bandit cut the aggregate cold-start rate from 63\%
under immediate scale-to-zero to 3\%, held tail (P95) latency at the level of the
longest-retention fixed baseline, and used roughly 13\% less accumulated warm time
than that baseline, while also beating the predictive baseline on cold starts and
latency. No fixed timeout was best on every metric, whereas the adaptive agents
reached a favorable operating point without manual tuning; on the bursty workload at a
converged horizon, the bandit's advantage over the reactive and predictive baselines is
statistically significant. On a multi-node Kubernetes cluster under the platform's own
Horizontal Pod Autoscaler, the controller coexists with an autoscaler it does not
control: the replica count follows load identically across policies, the replicas spread
across nodes, and retention independently governs warm time.
\end{abstract}

\begin{IEEEkeywords}
serverless computing, Function-as-a-Service, cold start, warm-state retention,
keep-alive, multi-armed bandit, AIMD, online control, OpenFaaS.
\end{IEEEkeywords}

\section{Introduction}
Serverless computing has become a common way to build event-driven and cloud-native
applications, most often through the Function-as-a-Service (FaaS)
model~\cite{shafiei2022serverless,baldini2017serverless,taibi2021serverless}.
Developers write small, stateless functions, and the platform takes care of
deploying, scaling, and running them, paying for compute only while a request is
being handled. This model is convenient, but it incurs a recurring cost. When a
function is invoked and no initialized environment is available, the platform must
first provision and initialize one. This delay, the \emph{cold start}, arises from
several steps in sequence: allocating resources, starting the runtime, loading
dependencies, and performing application-level setup~\cite{golec2024coldstart}.

Cold starts are not a rare corner case. Studies of production traffic report that
invocation patterns are highly uneven, with a long tail of functions that are
invoked only occasionally and remain idle in between~\cite{shahrad2020serverless}.
Recent end-to-end measurements show that cold starts occur frequently enough to
affect tail latency measurably~\cite{joosen2025coldstarts}. Because idle
environments are reclaimed first, it is precisely these irregular functions that
repeatedly incur the reinitialization cost.

The usual defense is \emph{warm-state retention}, or keep-alive: hold an idle
environment for a bounded period so that a subsequent request can reuse it before it
is torn down. In practice this window is almost always a \emph{static}
setting, such as a fixed idle timeout or a minimum replica count~\cite{akkus2018sand}.
This is simple to configure but difficult to tune. A window suited to one
function's traffic is either wasteful or too short for another, and a value chosen
for current conditions degrades once the workload shifts. Any fixed choice pins the
operator to a single point on a trade-off curve: a short window conserves resources
but admits more cold starts, while a long window suppresses cold starts but pays for
idle capacity.

Our premise is that this window need not be a fixed constant. We treat the
keep-alive time of a function~$f$ as a runtime parameter $T_{\mathrm{warm}}(f)$, the
time at least one environment remains available after the last invocation, and we
adjust it \emph{online}: every update is computed during execution from the outcomes
observed so far, with no offline training phase and no knowledge of future
invocations. One design principle is central to the paper: we deliberately separate
the retention question from autoscaling, scoping this study to the single-instance
regime where retention duration is the dominant lever. Most adaptive serverless work asks \emph{how many} replicas to run; we
address the narrower and, we argue, more specific question of \emph{how long} a
single environment should remain available after it becomes idle, the decision
that dominates when one warm instance suffices.

We make the following contributions:
\begin{itemize}
\item We recast per-function keep-alive as an online control variable decoupled from
autoscaling, framing retention \emph{duration} (not replica count) as an explicit,
lightweight, training-free control target rather than a fixed timeout
(Sec.~\ref{sec:design}).
\item We design and implement an \emph{external} retention controller for
OpenFaaS/faasd that enforces $T_{\mathrm{warm}}$ without modifying the platform,
driven by two agents, a reactive AIMD rule and a UCB1 bandit, that need no
training, no workload model, and only $O(\#\text{functions})$ state, so the approach
ports to any FaaS runtime (Secs.~\ref{sec:design}--\ref{sec:impl}).
\item We evaluate both agents against three static baselines and a predictive
histogram keep-alive on steady, bursty, and trace-derived workloads (cold-start rate,
mean and P95 latency, warm time), over 10{,}800 logged invocations, plus a sensitivity
sweep over the agents' parameters (Secs.~\ref{sec:setup}--\ref{sec:results}). The
bandit attains cold-start and tail-latency behavior close to the longest-retention
baseline at $\sim$13\% less warm time and, despite being model-free, outperforms the
predictive histogram baseline; no single configuration is best on every metric. We
further establish the bandit's advantage statistically at a converged horizon
(Sec.~\ref{sec:signif}), show the policy ordering carries over to two additional realistic
functions, a JSON API and image resizing (Sec.~\ref{sec:morefns}), and, on a multi-node
Kubernetes cluster under the stock Horizontal Pod Autoscaler, show the controller coexists
with an autoscaler it does not control, whose replica count follows load and spans multiple
nodes independently of the retention policy (Sec.~\ref{sec:multireplica}).
\end{itemize}

\section{Background and Related Work}
\label{sec:related}

\subsection{Cold Starts and Warm-State Retention}
Every FaaS invocation runs inside an execution environment (a container or
micro-VM). A \emph{warm} call reuses an initialized one and skips startup; a
\emph{cold} call must build one first~\cite{golec2024coldstart}. Longer retention cuts
cold starts but holds idle resources longer, so the best timeout depends on traffic:
a busy function benefits from a long window, while a rarely called one gains little
and mostly ties up capacity~\cite{shahrad2020serverless}.

\subsection{Heuristic and Cache-Based Retention}
Early systems such as SAND improve performance by increasing environment reuse and
reducing initialization overhead~\cite{akkus2018sand}. FaasCache reframes keep-alive
as a caching problem, applying a Greedy-Dual policy that prioritizes which
containers to retain under resource pressure based on recency and startup
cost~\cite{fuerst2021faascache}. PULSE reduces keep-alive cost using mixed-quality
predictive models to guide retention~\cite{sankaranarayanan2024pulse}. These
approaches largely decide \emph{which} environments to keep, using predefined rules
or prioritization logic, and adapt retention only coarsely at runtime; the
\emph{temporal} question of how long a given function should stay warm receives
less attention.

\subsection{Forecast-Based Retention}
A second line predicts future demand and pre-warms accordingly. Deep-learning
frameworks model invocation behavior as time series to drive cold-start
management~\cite{nguyen2024holistic}, and systems such as IceBreaker forecast when
each function will next be invoked and warm it in advance, selecting lower-cost
hardware during the wait~\cite{roy2022icebreaker}. These methods are effective when
traffic is predictable, but their performance depends on forecast accuracy;
irregular, externally driven workloads are difficult to predict, and a prediction
error causes either wasted retention or premature teardown. They also require
historical data and periodic retraining.

\subsection{Reinforcement Learning and Online Adaptation}
Reinforcement learning (RL) treats resource management as sequential decision
making and can outperform static heuristics~\cite{agarwal2024rlserverless}, but most
RL formulations target \emph{autoscaling} (replica counts), fold retention in only
indirectly, and need richer state and more interaction data to stabilize. Lightweight
online methods are a middle ground: a multi-armed bandit selects among actions and
learns their payoffs with minimal state~\cite{lattimore2020bandit}, and UCB1 gives
finite-time regret guarantees while balancing exploration and
exploitation~\cite{auer2002ucb}; AIMD~\cite{chiu1989aimd} is an even simpler reactive
rule.

\subsection{Positioning}
Most existing work on cold starts is heuristic, forecast-based, or RL-based, and
across all three the keep-alive window is seldom treated as an explicit
per-function knob. The attention goes to \emph{how many} environments to keep, not
\emph{how long} to keep one, even though that timing is the primary factor behind
cold starts under irregular and bursty traffic. Our work addresses this gap. We make
retention duration the variable of interest, adapt it online with two intentionally
simple agents (AIMD and UCB1), and study its effect in isolation from autoscaling,
later showing the two loops compose (Sec.~\ref{sec:multireplica}). Table~\ref{tab:compare} positions our approach against representative prior
work on the dimensions most relevant to runtime retention: runtime adaptation,
freedom from a training/forecast phase, direct control of retention \emph{duration},
and computational lightness.

\begin{table}[t]
\caption{Positioning against representative prior work. \emph{part.}\ marks
mechanisms that affect retention \emph{duration} only indirectly.}
\label{tab:compare}
\centering
\renewcommand{\arraystretch}{1.2}
\setlength{\tabcolsep}{3.2pt}
\begin{tabular}{l ccccc}
\toprule
\textbf{Approach} & \textbf{Runtime} & \textbf{No} & \textbf{Controls} & \textbf{Light-}\\
 & \textbf{adapt.} & \textbf{train} & \textbf{duration} & \textbf{weight}\\
\midrule
SAND~\cite{akkus2018sand}            & --   & \checkmark & --        & \checkmark\\
FaasCache~\cite{fuerst2021faascache} & part.& \checkmark & part.     & \checkmark\\
PULSE~\cite{sankaranarayanan2024pulse}& \checkmark & -- & part.     & --\\
Forecast~\cite{nguyen2024holistic,roy2022icebreaker} & \checkmark & -- & part. & --\\
RL~\cite{agarwal2024rlserverless}    & \checkmark & -- & part.     & --\\
\textbf{This work}                   & \checkmark & \checkmark & \checkmark & \checkmark\\
\bottomrule
\end{tabular}
\end{table}

\section{System Design}
\label{sec:design}

\subsection{Problem Formulation}
Let a function $f$ have a retention duration $T_{\mathrm{warm}}(f)$ specifying how
long at least one execution environment remains available after its most recent
invocation. A static policy fixes $T_{\mathrm{warm}}$ for the entire run; an
adaptive policy updates it during execution from observed outcomes. The objective
is to dynamically choose $T_{\mathrm{warm}}$ so as to reduce cold-start frequency
and latency while avoiding unnecessary accumulated warm time, under workloads whose
invocation behavior is not known in advance.

\subsection{Requirements}
The system must (R1) adjust $T_{\mathrm{warm}}$ at runtime; (R2) operate at
per-function granularity, since invocation behavior differs across functions;
(R3) support multiple retention strategies (static and adaptive) under identical
conditions for fair comparison; (R4) produce reproducible warm/cold transitions;
(R5) exercise heterogeneous workloads; and, on the non-functional side, (R6) remain
lightweight, (R7) stay modular by separating retention \emph{enforcement} from policy
\emph{decision}, and (R8) work with an unmodified serverless platform.

\subsection{Architecture}
\begin{figure}[t]
\centering
\includegraphics[width=\columnwidth]{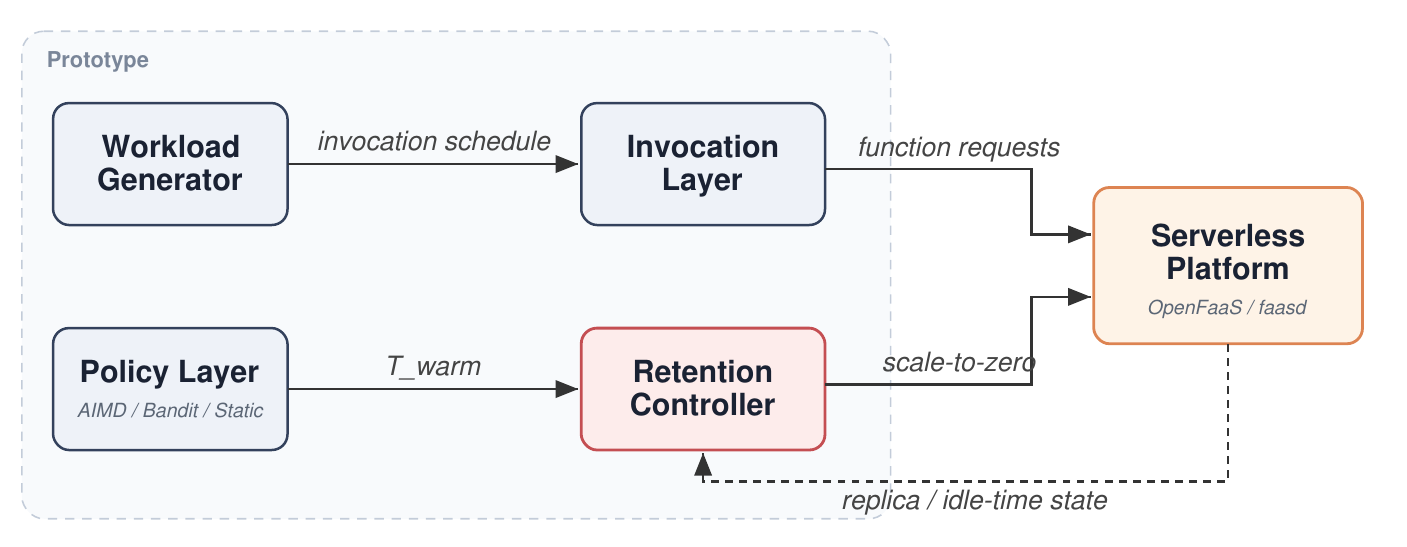}
\caption{System architecture. The retention controller enforces the per-function
control parameter $T_{\mathrm{warm}}$ supplied by the policy layer, using only the
platform's external interfaces (replica/idle state in, scale-to-zero out).}
\label{fig:arch}
\end{figure}

Fig.~\ref{fig:arch} shows the four components. A \emph{workload generator} emits
invocation sequences; an \emph{invocation layer} forwards them to the OpenFaaS
gateway and records latency and platform state; a \emph{policy layer} computes the
per-function $T_{\mathrm{warm}}$; and a \emph{retention controller} enforces it by
observing replica/idle state and issuing scale-to-zero. Decoupling enforcement
(controller) from decision (policy) satisfies R3 and R7: any policy plugs in
without changing lifecycle logic.

\subsection{Warm-State Lifecycle}
\begin{figure}[t]
\centering
\includegraphics[width=\columnwidth]{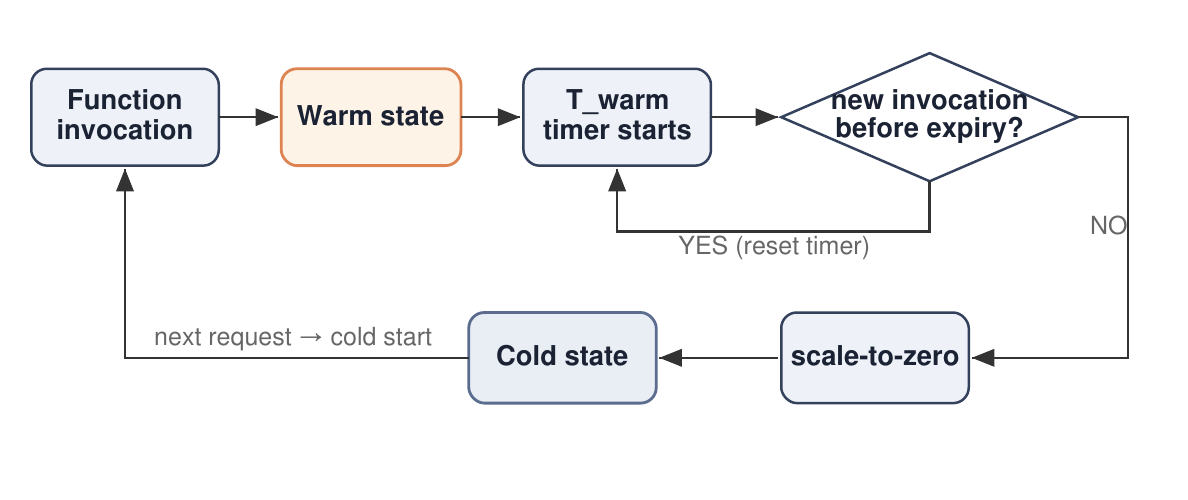}
\caption{Simplified binary warm/cold lifecycle. Retention duration
$T_{\mathrm{warm}}$ governs the warm$\rightarrow$cold transition; it is the single
control variable adapted at runtime.}
\label{fig:life}
\end{figure}

To isolate retention from autoscaling, we adopt a binary lifecycle (Fig.~\ref{fig:life}):
a function is either \emph{warm} (exactly one available environment) or \emph{cold}
(none). An invocation makes the function warm and starts a retention timer of length
$T_{\mathrm{warm}}$. A request arriving before expiry is served warm and resets the
timer; otherwise the controller scales the function to zero, and the next request
incurs a cold start. Restricting to one warm instance improves interpretability and
removes replica-scaling effects, so observed differences are attributable to
retention decisions.

\subsection{Retention Strategies}
We evaluate three static baselines, two adaptive agents, and a predictive baseline.

\paragraph{Static baselines}
$T_{\mathrm{warm}}\in\{0,10,60\}$\,s. The value $0$ scales to zero immediately on
idle (shortest retention); $60$ retains the longest; $10$ is an intermediate
setting.

\paragraph{AIMD agent}
A reactive feedback rule inspired by congestion control~\cite{chiu1989aimd}.
A cold execution multiplicatively increases retention to suppress future cold
starts; a sustained run of warm executions additively decreases it to release idle
capacity:
\begin{equation}
T_{\mathrm{warm}} \leftarrow
\begin{cases}
\min(a\cdot T_{\mathrm{warm}},\,T_{\max}) & \text{on a cold execution},\\[2pt]
\max(T_{\mathrm{warm}}-b,\,T_{\min}) & \text{after a warm-streak},
\end{cases}
\end{equation}
with multiplicative factor $a$, additive step $b$, and bounds
$[T_{\min},T_{\max}]$ that prevent unstable adaptation. The rule reacts strongly to
cold starts and relaxes gradually during stable warm phases.

\paragraph{Bandit (UCB1) agent}
An online-learning policy that selects $T_{\mathrm{warm}}$ from a discrete action
set $\mathcal{A}=\{10,20,60\}$\,s, treating each duration as an arm. After a
round-robin warm-up that samples each arm once, it chooses by the UCB1
rule~\cite{auer2002ucb}:
\begin{equation}
a_t = \arg\max_{i\in\mathcal{A}}
\left(\hat{\mu}_i + \sqrt{\tfrac{2\ln t}{n_i}}\right),
\end{equation}
where $\hat{\mu}_i$ is the mean reward of arm $i$ and $n_i$ its selection count.
The reward penalizes both sides of the trade-off, charging a fixed penalty for a cold
execution and a smaller penalty that grows with the chosen retention duration, so the
agent gravitates toward windows that keep both cold starts and warm time down. It
needs no workload forecast and no offline training.

\paragraph{Predictive baseline (histogram)}
As an adaptive reference, we add a Shahrad-style predictive keep-alive~\cite{shahrad2020serverless}:
per function it keeps a short history of recent inter-arrival times and sets
$T_{\mathrm{warm}}$ to their 95th percentile, so regular traffic gets a short window
and bursty traffic a longer one. Like the agents it adapts online, but it estimates
the arrival distribution rather than optimizing the trade-off directly.

\subsection{Retention Controller}
\begin{figure}[t]
\centering
\includegraphics[width=0.86\columnwidth]{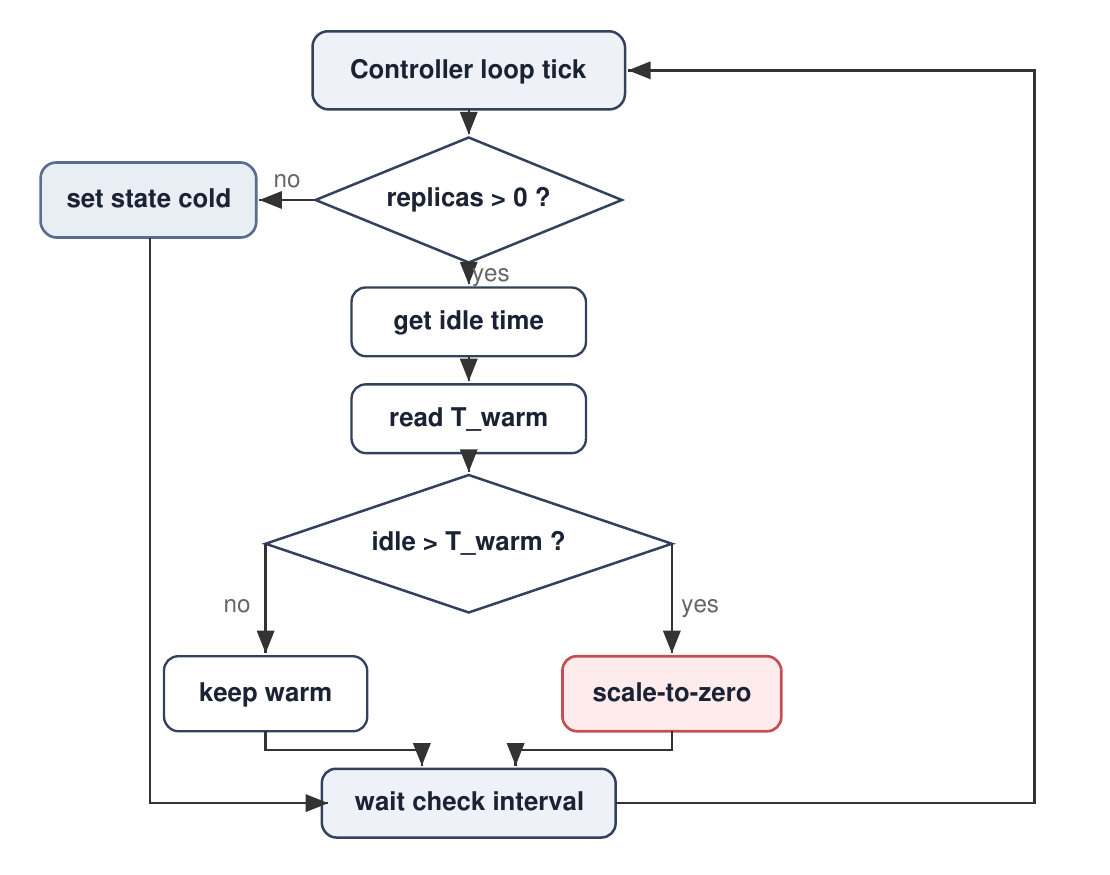}
\caption{Retention controller loop. Policy decisions enter only through the shared
$T_{\mathrm{warm}}$ store; lifecycle enforcement (scale-to-zero) is fully contained
in the controller.}
\label{fig:ctrl}
\end{figure}

Because OpenFaaS/faasd exposes no configurable post-idle keep-alive, a dedicated
controller (Fig.~\ref{fig:ctrl}) supplies it externally. On each tick it checks
whether a function has active replicas; if not, it records the cold state. If warm,
it reads the function's idle time and the current $T_{\mathrm{warm}}$ from a shared
store and, when idle time exceeds $T_{\mathrm{warm}}$, triggers scale-to-zero;
otherwise the function is kept warm. Adaptive policies write updated durations to
the shared store, while enforcement stays entirely within the controller (R7).

Algorithm~1 states the controller loop together with the two
policy update rules. Enforcement (lines 1--8) is policy-agnostic and reads only
the shared $T_{\mathrm{warm}}$ store; the AIMD and bandit update routines write to
that store after each invocation, so a new policy can be added without touching
lifecycle logic.

\begin{figure}[t]
\centering
\setlength{\fboxsep}{0pt}
\noindent\rule{\columnwidth}{0.9pt}\par\vspace{1pt}
\noindent\textbf{Algorithm 1:} Retention enforcement and policy updates\par
\vspace{1pt}\noindent\rule{\columnwidth}{0.4pt}\par\vspace{2pt}
{\footnotesize\ttfamily\setlength{\parindent}{0pt}\raggedright
\textbf{Controller loop} (every $\Delta$ seconds):\\
\makebox[1.8em][l]{1}\textbf{for} each monitored function $f$:\\
\makebox[1.8em][l]{2}\hspace*{1.2em}\textbf{if} replicas$(f)=0$: state$(f)\!\leftarrow$ cold; \textbf{continue}\\
\makebox[1.8em][l]{3}\hspace*{1.2em}idle $\leftarrow$ idleTime$(f)$;\, $T \leftarrow$ store$[f]$\\
\makebox[1.8em][l]{4}\hspace*{1.2em}\textbf{if} idle $> T$:\\
\makebox[1.8em][l]{5}\hspace*{2.4em}scaleToZero$(f)$; state$(f)\!\leftarrow$ cold\\
\makebox[1.8em][l]{6}\hspace*{1.2em}\textbf{else}: keepWarm$(f)$\\[3pt]
\textbf{AIMD update} (after invocation, outcome $o$):\\
\makebox[1.8em][l]{7}\textbf{if} $o{=}$cold: $T{\leftarrow}\min(aT,T_{\max})$; streak${\leftarrow}0$\\
\makebox[1.8em][l]{8}\textbf{else if} streak${\ge}S$: $T{\leftarrow}\max(T{-}b,T_{\min})$; streak${\leftarrow}0$\\[3pt]
\textbf{Bandit (UCB1) update} (after invocation):\\
\makebox[1.8em][l]{9}$r \leftarrow -(20\!\cdot\![o{=}\text{cold}]) - 0.05\,T$\\
\makebox[1.8em][l]{10}update $\hat{\mu}_T,\,n_T$ with $r$;\, $t{+}{+}$\\
\makebox[1.8em][l]{11}store$[f] \leftarrow \arg\max_i\,\big(\hat{\mu}_i + \sqrt{2\ln t / n_i}\big)$\\
}
\vspace{1pt}\noindent\rule{\columnwidth}{0.9pt}
\end{figure}

\section{Implementation}
\label{sec:impl}
The prototype is implemented in Python~3.12 and runs on an Ubuntu~24.04 guest in a
lightweight VM on an Apple M1 Pro host (16\,GB RAM, macOS~26.5.1).
OpenFaaS is deployed with the lightweight \texttt{faasd} runtime over
\texttt{containerd} and \texttt{systemd}, chosen over Kubernetes to minimize
orchestration complexity while retaining well-defined deployment and invocation
interfaces and direct access to the container lifecycle, a practical compromise
between realism and controllability.

\paragraph{Functions}
Two benchmark functions use the \texttt{python3-http} template. \texttt{fastpy} is
a lightweight handler with minimal startup cost; \texttt{slowinitpy} adds an
initialization delay executed only on a cold path, modeling a more
cold-start-sensitive workload. Sharing the runtime and deployment configuration
isolates initialization behavior as the only difference.

\paragraph{Controller and scale-to-zero}
The controller runs as an external Python process with a configurable check
interval, monitored function set, and shared $T_{\mathrm{warm}}$ store. Since
\texttt{faasd} provides no native post-idle retention, scale-to-zero is realized by
interacting directly with \texttt{containerd}: tasks are terminated and deleted via
\texttt{ctr}, then containers and images removed, so the next invocation must
recreate the environment. A scheduling lock prevents concurrent scale operations on
a function and retry logic improves teardown reliability.

\paragraph{Policies}
The AIMD policy updates after each invocation, using multiplicative factor $a=2.0$
on a cold execution and additive decrease $b=2$\,s after five consecutive warm
executions (an error response resets the streak counter), with bounds
$T_{\mathrm{warm}}\in[10,60]$\,s. The bandit uses arms $\{10,20,60\}$\,s with a
cold-execution penalty of $20$ and a warm-cost penalty weighted by $0.05$ per
retained second, after a one-pass round-robin warm-up. The histogram baseline uses
the 95th percentile of the last 20 inter-arrival times, bounded to $[5,120]$\,s.

\paragraph{Cold-start detection}
\texttt{faasd} exposes no explicit cold-start signal, so the experiment runner queries
the active replica count immediately before each invocation and classifies an execution
as cold when $\mathrm{replicas}_{\text{before}}(f)=0$, a reproducible proxy for
the warm/cold path. We validated this proxy against logged deployment latency: every
cold-labeled invocation recorded a non-zero environment-creation latency (mean
0.2\,s) and every warm-labeled invocation recorded zero. This 0.2\,s is the
\texttt{containerd} container-recreation cost the controller measures; the multi-second
cold latencies reported later add application-level initialization on top (e.g.,
\texttt{slowinitpy}'s 3\,s), so the two are not in conflict. Transient connection failures
during initialization are handled by bounded retries. Every invocation logs platform state, cold classification,
latency components, the active $T_{\mathrm{warm}}$, and the policy decision.

\section{Experimental Setup}
\label{sec:setup}

\subsection{Workloads}
Three workload classes exercise different invocation dynamics:
\emph{steady} issues requests at a fixed 8\,s inter-arrival interval;
\emph{bursty} alternates bursts of five requests at 0.5\,s spacing with 25\,s idle
gaps; and \emph{trace-derived} replays the Huawei Serverless Data Release
2023~\cite{joosen2023huawei}. Because the trace provides per-second request counts
rather than replayable events, counts were expanded into individual invocation
events with relative timestamps (capped at 100 events, with idle gaps capped at
10\,min), and replayed with a $10\times$ speed-up to bound experiment duration while
preserving relative ordering.

\paragraph{Timescale}
All temporal parameters, namely inter-arrival intervals, idle gaps, retention
windows, and the trace replay, are deliberately compressed to a common laboratory
timescale. Because a function is served cold exactly when its idle gap exceeds the
retention window, cold-start behavior depends on the \emph{ratio} of window to idle
gap rather than on absolute durations. The baselines $\{0,10,60\}$\,s bracket the
gap scale of the workloads (0.5--25\,s): a window shorter than, comparable to, and
longer than the typical idle gap. Production keep-alive timers operate at minute
scale but face minute-to-hour idle gaps, placing them in the same three regimes.

\subsection{Configurations and Procedure}
Each configuration is a (policy, function, workload) triple. We evaluate six
policies (three static, AIMD, bandit, and the histogram baseline) $\times$ two
functions $\times$ three workloads. Every run executes 100 invocations and is
repeated three times under identical conditions (Table~\ref{tab:params}), with
function state reset before each run. In total,
$6\times2\times3\times3\times100=10{,}800$ invocations were logged, i.e., 300 per
(policy, function, workload) condition, on an Apple M1 Pro (16\,GB, macOS~26.5.1)
running an Ubuntu~24.04 VM.

\begin{table}[t]
\caption{Experimental Parameters}
\label{tab:params}
\centering
\renewcommand{\arraystretch}{1.2}
\begin{tabular}{ll}
\toprule
\textbf{Parameter} & \textbf{Configuration}\\
\midrule
Functions & \texttt{fastpy}, \texttt{slowinitpy}\\
Static policies & $T_{\mathrm{warm}}\in\{0,10,60\}$\,s\\
Adaptive policies & AIMD, Bandit (UCB1)\\
AIMD bounds / params & $[10,60]$\,s,\; $a{=}2.0$, $b{=}2$\,s, streak${=}5$\\
Bandit action space & $\{10,20,60\}$\,s\\
Predictive baseline & histogram, p95 of recent IATs, $[5,120]$\,s\\
Workloads & steady, bursty, trace-derived\\
Invocations per run & 100\\
Repetitions & 3\\
Platform & \texttt{faasd}, Ubuntu 24.04 VM (M1 Pro)\\
\bottomrule
\end{tabular}
\end{table}

\subsection{Metrics}
We report four metrics. The \emph{cold-start rate} is the fraction of invocations
classified cold (primary objective; lower is better). \emph{Mean response latency}
and the \emph{95th-percentile (P95) latency} capture average and tail responsiveness;
P95 is emphasized because cold executions create heavy-tailed delays. \emph{Accumulated warm time}
(the summed seconds environments stayed warm) stands in for retained
resource cost, where lower is better at a given cold-start level. Because retention
affects both responsiveness and resource use, no single metric is sufficient; the
relevant question is the trade-off each policy achieves. All reported values are
means over three repetitions; run-to-run variation is small (standard deviation
$\le 0.2$ percentage points for cold-start rate and $\le 5$\,ms for mean latency) and
is shown as error bars in Figs.~\ref{fig:cs}, \ref{fig:meanlat}, and~\ref{fig:p95}.

\section{Results and Discussion}
\label{sec:results}

\subsection{Cold-Start Reduction}
\begin{figure}[t]
\centering
\includegraphics[width=\columnwidth]{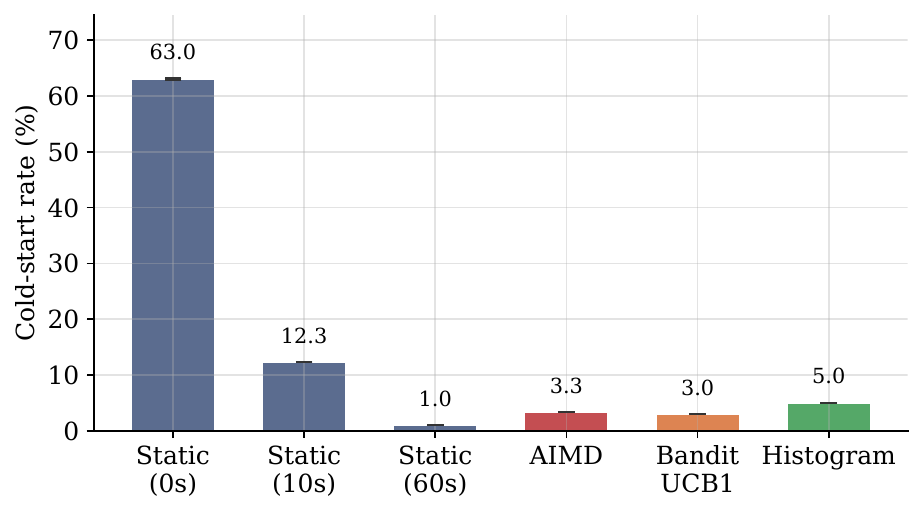}
\caption{Aggregate cold-start rate per policy (lower is better). The adaptive agents
(AIMD red, bandit orange) approach the 60\,s baseline and beat the predictive
histogram baseline (green). Error bars: std over three runs.}
\label{fig:cs}
\end{figure}

Fig.~\ref{fig:cs} reports the aggregate cold-start rates. Immediate scale-to-zero
(Static~0\,s) is the worst at 63.0\%, as expected, since every idle gap reclaims the
environment. Lengthening the fixed window reduces the rate substantially, to 12.3\%
at 10\,s and 1.0\% at 60\,s. Both adaptive agents reach the same low-cold-start
regime without a manually selected long timeout, at 3.3\% for AIMD and 3.0\% for the
bandit, and both beat the predictive histogram baseline (5.0\%).

\begin{figure}[t]
\centering
\includegraphics[width=\columnwidth]{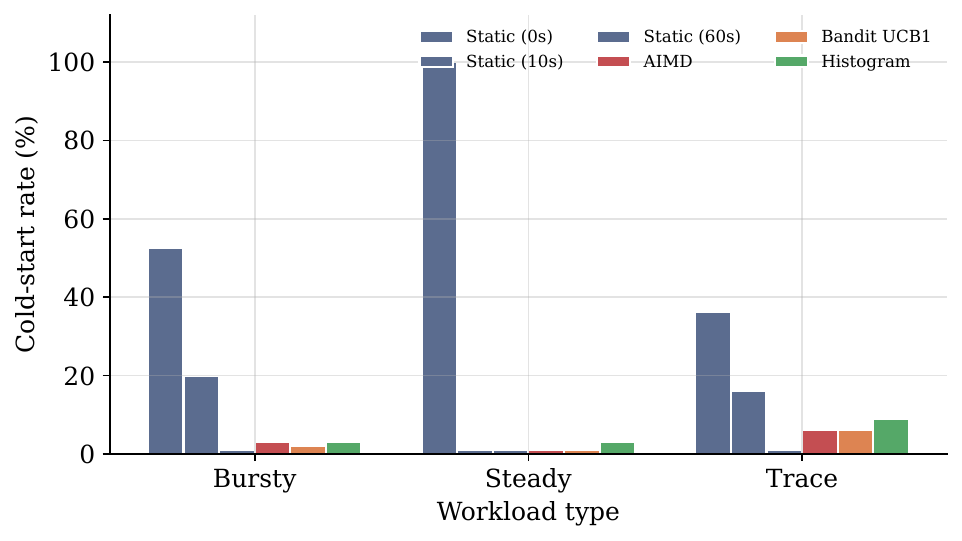}
\caption{Cold-start rate per workload and policy. The best static window depends on
the workload (e.g., Static~10\,s is good for steady but weak for bursty traffic),
whereas adaptive policies remain low throughout.}
\label{fig:csw}
\end{figure}

Breaking the same numbers down by workload (Fig.~\ref{fig:csw}) shows why a fixed
window fails to generalize across workloads. Static~0\,s swings from 36\% on the trace
to a full 100\% on steady traffic, where the 8\,s arrival interval always outlasts the
zero window and every request ends up cold. Static~10\,s flips the story: excellent on
steady traffic (1\%) but poor on bursty traffic (20\%), because the 25\,s idle gaps
outlive the window. No single fixed value covers all three patterns. The adaptive
agents stay at or below 6\% throughout, with the bandit reaching 1\% (steady) and 2\%
(bursty), and the histogram baseline stays within 3--9\%.

\subsection{Latency}
\begin{figure}[t]
\centering
\includegraphics[width=\columnwidth]{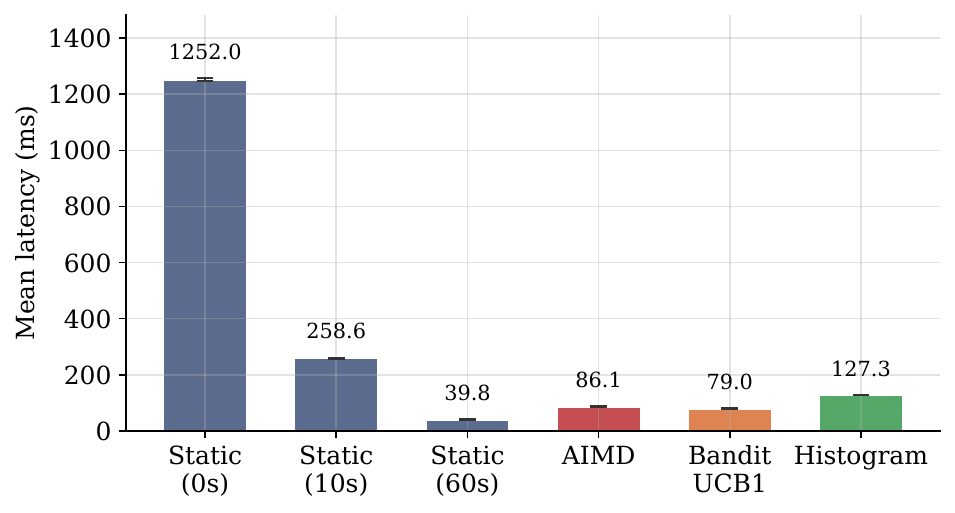}
\caption{Mean response latency per policy. Adaptive agents cut mean latency by
$\sim$94\% versus immediate scale-to-zero (they approach the longest-retention baseline
on the tail; mean latency remains above it).}
\label{fig:meanlat}
\end{figure}

\begin{figure}[t]
\centering
\includegraphics[width=\columnwidth]{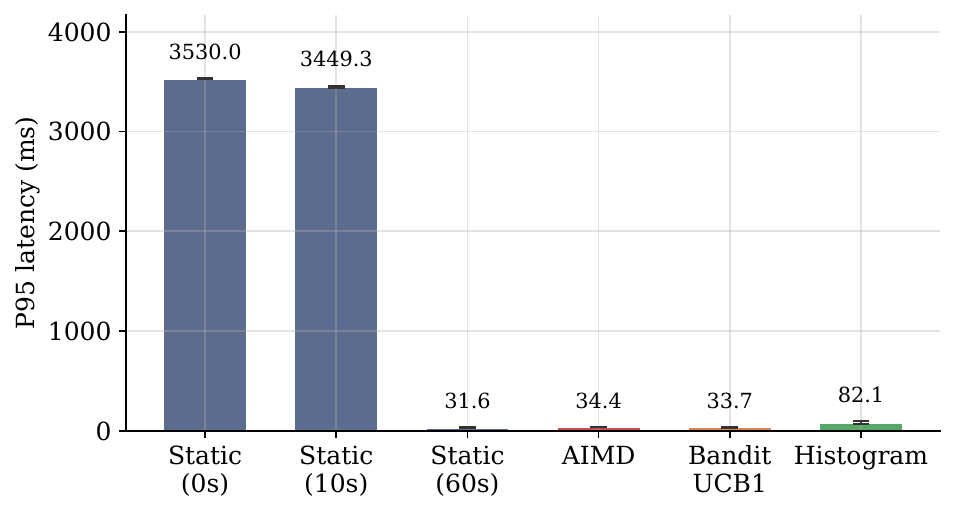}
\caption{P95 (tail) latency, pooled per policy. Short static windows incur multi-second
tails; Static~60\,s, AIMD, and Bandit cluster near 32--34~ms, while the histogram
baseline is higher (per-function tails in Table~\ref{tab:perfn}).}
\label{fig:p95}
\end{figure}

Because cold executions dominate latency, the latency results follow the cold-start
results (Figs.~\ref{fig:meanlat}--\ref{fig:p95}). Mean latency falls from 1252~ms at
Static~0\,s to 40~ms at Static~60\,s; the bandit reaches 79~ms and AIMD 86~ms, about
94\% below scale-to-zero, though still above Static-60's 40~ms, while the predictive
baseline sits higher at 127~ms. The parity with the fixed baseline is on the tail, not
the mean. Static~0\,s and Static~10\,s show P95 tails above 3400~ms owing to frequent
cold starts, whereas Static~60\,s, AIMD, and the bandit cluster tightly
(Fig.~\ref{fig:p95}); per function the bandit's tail matches the longest fixed baseline
exactly (22~ms on \texttt{fastpy}, 35~ms on \texttt{slowinitpy}, Table~\ref{tab:perfn}),
while the histogram is higher and function-dependent (82 and 213~ms). We report P95 per
function, since percentiles do not average across a mixed population. In effect, warm
executions occupy a similar low-latency band
across policies, and the policies differ mainly in how often a request incurs the
slow cold-start tail.

\subsection{Warm Time and the Trade-off}
\begin{figure}[t]
\centering
\includegraphics[width=\columnwidth]{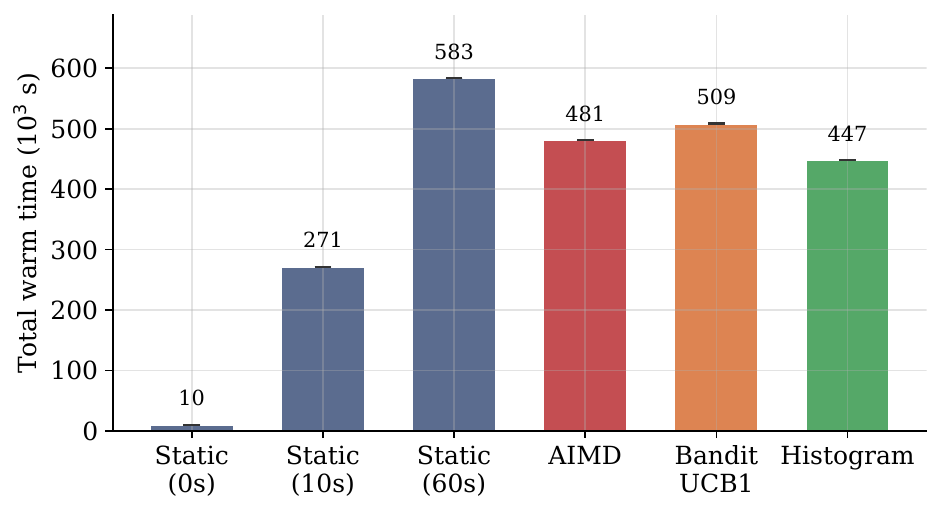}
\caption{Total accumulated warm time per policy (proxy for retained resource cost).
Both adaptive agents use less warm time than the longest-retention 60\,s baseline.}
\label{fig:warm}
\end{figure}

For the static policies, accumulated warm time (Fig.~\ref{fig:warm}) grows as
expected: 9.9$\times$10$^3$\,s at 0\,s, 272$\times$10$^3$\,s at 10\,s, and
583$\times$10$^3$\,s at 60\,s. The adaptive agents fall between the extremes, at
481$\times$10$^3$\,s (AIMD) and 509$\times$10$^3$\,s (bandit), both under the 60\,s
baseline. This is the central result: the bandit approaches Static~60\,s on cold
starts (3.0\% vs.\ 1.0\%) and matches it on tail latency, at roughly 13\% less warm
time. The predictive baseline uses even less warm time (447$\times$10$^3$\,s) but
pays with more cold starts (5.0\%), a different, less favorable point on the same
curve. In trade-off terms, each static timeout commits the operator to a single
cold-start/warm-time operating point chosen in advance, whereas the adaptive agents
settle at a favorable point, namely low cold starts at less warm time than the longest
baseline, without any prior knowledge of the workload.

\subsection{Adaptive Policy Behavior}
We next verify that the agents adapt to the workload rather than converging on a
single fixed value. Table~\ref{tab:twarm} reports the mean retention duration each
agent selects per workload, and the pattern is consistent with the traffic. On steady
traffic, where an 8\,s interval is covered by a short window, both agents settle low
(AIMD 11.5\,s, bandit 20.7\,s) and conserve warm time. On bursty traffic the 25\,s
idle gaps make early teardown costly, and both increase their windows (AIMD to
34.5\,s); the bandit rises further to 58.3\,s, near its longest arm, trading warm
time for cold-start avoidance.
The trace, with irregular spacing, yields AIMD $43.2$\,s and bandit $39.9$\,s; for the
bandit this sits between its steady and bursty settings, whereas AIMD, reacting to each
cold start, settles slightly higher than on bursty traffic. The two reach similar
outcomes by different mechanisms:
AIMD responds multiplicatively to each cold start, whereas the bandit's reward
estimates steer it toward the duration that best balances the cold-start penalty
against the warm-time penalty for that workload. The histogram baseline tracks the
arrival distribution in the same direction (8.5\,s on steady, 24.1\,s on bursty),
confirming that all three policies adapt per workload rather than settling on one
value.

\begin{table}[t]
\caption{Mean Selected $T_{\mathrm{warm}}$ (s) by Workload}
\label{tab:twarm}
\centering
\renewcommand{\arraystretch}{1.15}
\begin{tabular}{lccc}
\toprule
\textbf{Policy} & \textbf{Steady} & \textbf{Bursty} & \textbf{Trace}\\
\midrule
AIMD      & 11.5 & 34.5 & 43.2\\
Bandit    & 20.7 & 58.3 & 39.9\\
Histogram &  8.5 & 24.1 & 31.0\\
\bottomrule
\end{tabular}
\end{table}

\subsection{Effect of Function Startup Cost}
The two benchmark functions differ only in how expensive their initialization is, so
looking at them separately (Table~\ref{tab:perfn}) shows how startup cost changes the
payoff from retention control. The cold-start \emph{rate} is basically the same for
both functions under a given policy, which is expected: the rate is determined by
invocation timing, not by how heavy the init is. The latency consequences are a
different matter. A cold call to \texttt{slowinitpy} is much costlier, so Static~0\,s
runs to a 2240~ms mean and a 3565~ms P95, against 264~ms and 529~ms for
\texttt{fastpy}. Avoiding cold starts therefore yields a larger benefit on the
cold-start-sensitive function, and the adaptive agents capture it: the bandit reduces
\texttt{slowinitpy}'s mean latency from 2240~ms to 129~ms, a 94\% reduction, and
matches the longest-retention baseline on P95 (35~ms for both). The same
per-function control thus provides the greatest benefit where cold starts are most
costly, which motivates making $T_{\mathrm{warm}}$ per-function (R2).

\begin{table}[t]
\caption{Per-Function Results (Aggregated over Workloads)}
\label{tab:perfn}
\centering
\renewcommand{\arraystretch}{1.15}
\setlength{\tabcolsep}{4.5pt}
\begin{tabular}{l rrr rrr}
\toprule
& \multicolumn{3}{c}{\texttt{fastpy}} & \multicolumn{3}{c}{\texttt{slowinitpy}}\\
\cmidrule(lr){2-4}\cmidrule(lr){5-7}
\textbf{Policy} & Cold\% & Mean & P95 & Cold\% & Mean & P95\\
& & (ms) & (ms) & & (ms) & (ms)\\
\midrule
Static 0\,s  & 61.7 & 264 & 529 & 64.3 & 2240 & 3565\\
Static 10\,s & 12.3 &  63 & 382 & 12.3 &  454 & 3507\\
Static 60\,s & 1.0  &  20 &  22 & 1.0  &   60 &   35\\
AIMD         & 3.3  &  30 &  28 & 3.3  &  142 &   36\\
Bandit       & 3.0  &  29 &  22 & 3.0  &  129 &   35\\
Histogram    & 5.0  &  51 &  82 & 5.0  &  204 &  213\\
\bottomrule
\end{tabular}
\end{table}

\subsection{Discussion}
Three observations stand out. First, the bandit's advantage over AIMD is
consistent with their mechanisms: across the three runs its cold-start rate
(3.0\% vs.\ 3.3\%) and mean latency (79 vs.\ 86\,ms) were lower. At this short horizon
the gap is within run-to-run variation; a longer-horizon study (Sec.~\ref{sec:signif})
isolates it and shows it is statistically significant. On tail latency the two are
effectively tied. The
bandit keeps a reward estimate per arm and balances exploration against exploitation,
so it converges on the arm suited to the current traffic; AIMD reacts more slowly to
pattern changes. Second, against the predictive histogram baseline the bandit is
clearly better on cold starts (3.0\% vs.\ 5.0\%) and mean latency (79 vs.\ 127\,ms);
the histogram trades those for slightly less warm time, so the lightweight online
agents are competitive with, and here better than, a predictive keep-alive without
ever estimating the arrival distribution. Third, at the 100-invocation horizon the two
agents are close because the online learner has little room to converge; the
longer-horizon analysis of Sec.~\ref{sec:signif} confirms the bandit's advantage widens
and becomes statistically significant once it does.

\subsection{Parameter Sensitivity}
To check robustness to the hand-set parameters, we swept them one at a time on the
trace workload with \texttt{slowinitpy} (Table~\ref{tab:sweep}). Behavior moves
smoothly and predictably: raising the bandit's cold-start penalty (10$\to$40) trades
warm time for fewer cold starts (10\%$\to$3\%), while the warm-cost weight does the
reverse (0.10 gives 12\%); a finer arm set and a gentler AIMD factor land in between.
The two lowest-cold-start settings (penalty $40$ and weight $0.025$) coincide because
both drive the bandit to the same longest-arm operating point. No setting collapses,
and the paper's defaults sit in the middle of the useful range, so the result is not an
artifact of a single lucky configuration.

\begin{table}[t]
\caption{Parameter sweep (trace workload, \texttt{slowinitpy}). Defaults in bold.}
\label{tab:sweep}
\centering
\renewcommand{\arraystretch}{1.12}
\setlength{\tabcolsep}{4pt}
\begin{tabular}{lccc}
\toprule
\textbf{Configuration} & \textbf{Cold\%} & \textbf{Mean (ms)} & \textbf{Warm (10$^3$s)}\\
\midrule
Bandit penalty 10        & 10.0 & 381 & 35.8\\
\textbf{Bandit penalty 20} & \textbf{6.0} & \textbf{236} & \textbf{65.4}\\
Bandit penalty 40        &  3.0 & 131 & 83.1\\
Bandit weight 0.025      &  3.0 & 131 & 83.1\\
Bandit weight 0.10       & 12.0 & 448 & 23.4\\
Bandit arms 5/15/30/60   &  7.0 & 272 & 56.2\\
AIMD $a{=}1.5$           &  8.0 & 307 & 45.1\\
AIMD $a{=}3.0$           &  4.0 & 167 & 75.2\\
\bottomrule
\end{tabular}
\end{table}

\subsection{Convergence and Statistical Significance}
\label{sec:signif}
The runs above use a 100-invocation horizon and three repetitions. To test whether the
bandit's edge is real once the online learner converges, we reran the \emph{bursty}
workload at a longer horizon (250 invocations) over eight repetitions, adding $\pm15\%$
jitter to inter-arrival times so repetitions genuinely differ (a per-repetition seed
keeps the policies paired). Table~\ref{tab:signif} gives per-policy means and pairwise
Mann-Whitney tests against the bandit. At the longer horizon the bandit converges to a
stable operating point and \emph{significantly} outperforms the baselines it is most
naturally compared to: versus AIMD it lowers the cold-start rate from $2.3\%$ to
$0.8\%$ ($p{=}0.0003$) and mean latency from $76.8$ to $45.4$\,ms ($p{=}0.0008$); versus
the histogram predictor the margins are larger, including tail latency
($33$ vs.\ $647$\,ms). The longest fixed baseline, Static-60, edges the bandit on cold
starts ($0.4\%$ vs.\ $0.8\%$) only because a $60$\,s window happens to suit this
workload; the bandit reaches comparable behavior \emph{without} being given that
window, at slightly lower warm time than Static-60 ($2592$ vs.\ $2641$ replica-seconds,
$p{=}0.005$, a small margin; AIMD and the histogram in fact use less warm time still).
The bandit and
Static-60 are deterministic across repetitions (they settle clear of the idle-gap
boundary), so the significance is carried by the run-to-run spread of AIMD and the
histogram, whose distributions do not overlap the bandit's operating point.

\begin{table}[t]
\caption{Convergence/significance study (bursty workload, $250$ invocations,
$8$ jittered repetitions). Per-policy means; markers give the two-sided Mann-Whitney
significance of each baseline \emph{vs.\ the bandit} for that metric:
$^{***}p{<}0.001$, $^{**}p{<}0.01$, $^{*}p{<}0.05$, \textit{ns}~not significant
($n{=}8$). AIMD and the histogram carry run-to-run spread (e.g., AIMD cold-rate
$95\%$ CI $[2.2,2.4]$); the bandit and Static-60 are deterministic across repetitions,
so significance rests on the baselines' spread not overlapping the bandit's value.}
\label{tab:signif}
\centering
\renewcommand{\arraystretch}{1.15}
\setlength{\tabcolsep}{5pt}
\begin{tabular}{lcccc}
\toprule
\textbf{Policy} & \textbf{Cold\%} & \textbf{Mean (ms)} & \textbf{P95 (ms)} & \textbf{Warm (s)}\\
\midrule
Static-60       & $0.4^{***}$ & $36.3^{***}$ & $31.6^{*}$ & $2641^{**}$\\
\textbf{Bandit} & \textbf{0.8} & \textbf{45.4} & \textbf{33.1} & \textbf{2592}\\
AIMD            & $2.3^{***}$ & $76.8^{***}$ & $34.7^{\,\textit{ns}}$ & $2392^{***}$\\
Histogram       & $6.1^{***}$ & $154.8^{***}$ & $646.7^{***}$ & $1880^{***}$\\
\bottomrule
\end{tabular}
\end{table}

\subsection{Generality Across Function Types}
\label{sec:morefns}
To check that the results are not specific to the two original functions, we evaluated
every policy on two \emph{additional, more realistic} functions: a JSON transform API
(parse, validate, and aggregate a record set) and an image-resize function (decode and
produce several thumbnails with Pillow), run alongside the two originals. This ran as a
self-contained sweep on a single x86 \texttt{faasd} host (four functions $\times$ three
workloads $\times$ four policies $\times$ three repetitions); because it is a separate
testbed from the main \texttt{arm64} study we report it as its own comparison rather than
merging it into the main tables.

The policy ordering carries over unchanged to the realistic functions
(Table~\ref{tab:morefns}). Static-60 again attains the fewest cold starts but at the most
warm time; the bandit uses about $21\%$ less warm time than Static-60 ($523$ vs.\ $660$
$\times 10^3$ replica-seconds) at a modest cold-start cost, and it again beats the
predictive histogram on \emph{both} cold starts ($4.1\%$ vs.\ $6.8\%$) and, sharply, on
tail latency ($413$ vs.\ $2205$\,ms; the histogram's mispredictions inflate the tail).
Per-function absolute latency tracks each function's compute weight, from about $17$\,ms
warm for \texttt{fastpy} to about $430$\,ms for image resizing, while the relative ordering
of policies is the same for every function. This sweep used a fixed inter-arrival schedule
(no jitter), so it is a consistency check across function types rather than a significance
test; the significance result is established separately in Sec.~\ref{sec:signif}.

\begin{table}[t]
\caption{Generality across four functions (JSON API and image-resize added to the two
originals) on an x86 \texttt{faasd} testbed: per-policy means over three workloads and
three repetitions. The policy ordering is unchanged from the main study, and the bandit
beats the predictive histogram on both cold starts and tail latency
(Sec.~\ref{sec:morefns}).}
\label{tab:morefns}
\centering
\renewcommand{\arraystretch}{1.15}
\setlength{\tabcolsep}{5pt}
\begin{tabular}{lcccc}
\toprule
\textbf{Policy} & \textbf{Cold \%} & \textbf{Mean\,ms} & \textbf{P95\,ms} & \textbf{Warm ($10^3$s)}\\
\midrule
Static-60 & $1.4$ & $149$ & $396$ & $660$\\
AIMD & $4.5$ & $245$ & $420$ & $479$\\
Bandit & $4.1$ & $231$ & $413$ & $523$\\
Histogram & $6.8$ & $311$ & $2205$ & $420$\\
\bottomrule
\end{tabular}
\end{table}

\subsection{Decoupling from Autoscaling}
\label{sec:multireplica}
The main study fixes the replica count at one to isolate retention; \texttt{faasd}
cannot demonstrate coexistence with an autoscaler because it does not scale a function
horizontally. We therefore reproduce the decoupling on a genuine \emph{multi-node}
Kubernetes cluster (three-node \texttt{k3s}: one control-plane and two workers) under
the platform's own autoscaler. Crucially, horizontal scaling here is owned by the stock
Kubernetes \emph{Horizontal Pod Autoscaler} (HPA, \texttt{autoscaling/v2}, CPU-utilisation
target $50\%$, fed by the cluster metrics-server), an autoscaler we do not author or
control. Our retention controller acts only as the activation layer: once idle time
exceeds $T_{\mathrm{warm}}$ it removes the HPA and scales the Deployment to zero, and on
the next burst it restores the HPA and lets Kubernetes scale out again, after which the
first request is served by a freshly started replica (a genuine cold start). This
activation/scaling split mirrors how production autoscalers such as KEDA separate the
decision to keep a function alive from the decision of how many replicas to run. The
function is a CPU-bound HTTP workload so the CPU-driven HPA has a real signal to react
to. We sweep three idle-gap regimes ($18/45/90$\,s) $\times$ three policies $\times$ five
repetitions, ramping the offered load $1\!\to\!5\!\to\!1$ (maximum five replicas).

\begin{figure}[t]
\centering
\includegraphics[width=\columnwidth]{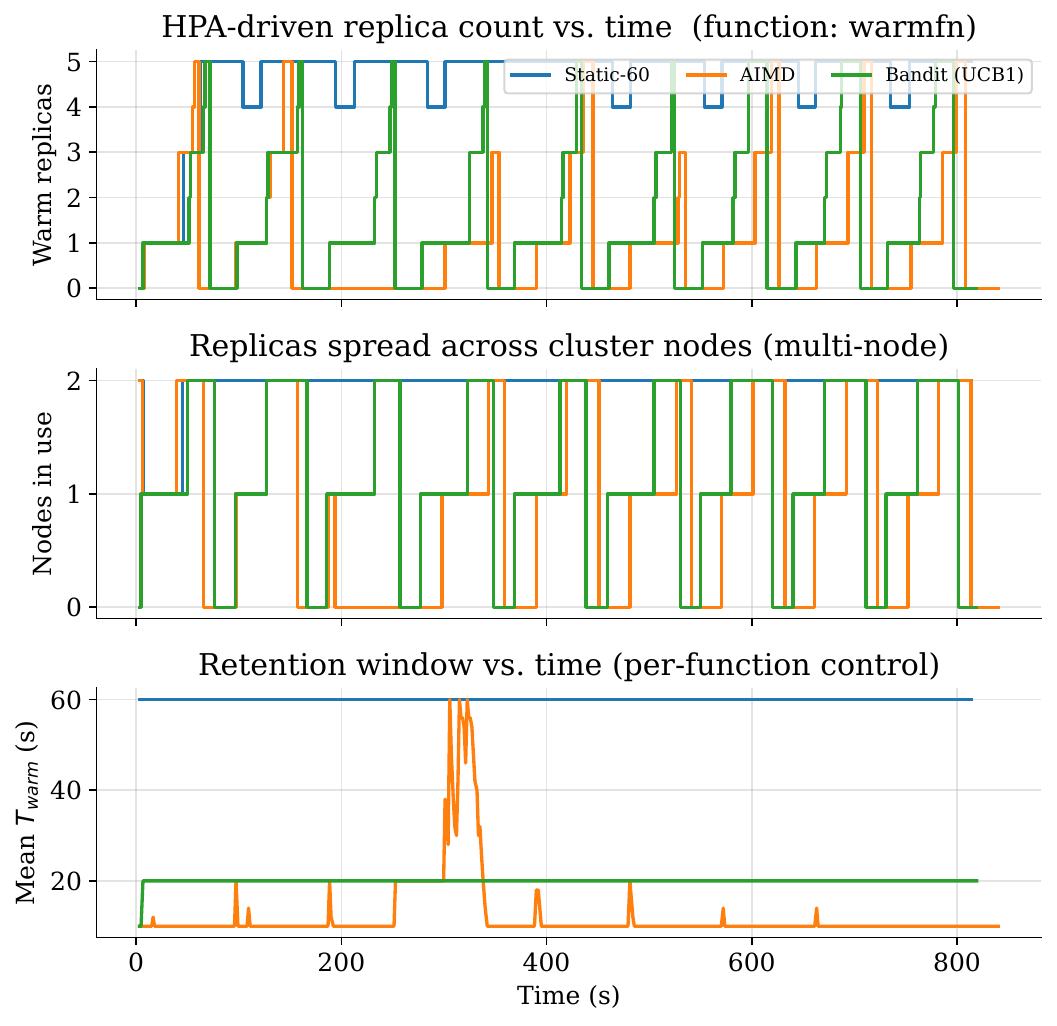}
\caption{Decoupling on a multi-node cluster under the stock Kubernetes HPA
($45$\,s idle gap). Top: the warm replica count follows the same load-driven
$1\!\to\!5$ trajectory for all policies (peak $5.0$, std $0.0$ across all cells).
Middle: replicas spread across both worker nodes (up to $2$). Bottom: the retention
window $T_{\mathrm{warm}}$ is set independently per policy (Static-60 fixed, bandit low,
AIMD adapting). The HPA sets the replica count from load; retention sets the warm floor
between bursts.}
\label{fig:mr}
\end{figure}

The replica-count side decouples cleanly, now against an autoscaler we did not write.
Across \emph{every} policy and regime the peak replica count is $5.0$ with standard
deviation $0.0$, and the replicas spread across both worker nodes in every cell (distinct
nodes in use $=2.0$, std $0.0$): under identical load the HPA drives every policy to the
same peak across the same two machines, regardless of which retention policy runs
(Fig.~\ref{fig:mr}, top and middle). What differs between policies is only how many
replicas persist \emph{between} bursts, which is precisely the retention lever and the
source of the warm-time separation below. Because the scaler is now a stock,
independently authored controller rather than our own loop, and because it genuinely
spans multiple nodes, this is a stronger test of composition than a single-node
custom-scaler setup: the two loops compose without interference even though neither was
written to be aware of the other. Retention is the lever that moves \emph{warm time}
(Fig.~\ref{fig:mr}, bottom; Table~\ref{tab:mr}): the adaptive policies hold far less warm
replica-time than Static-60 at a comparable, uniformly low cold-start rate. At the
$90$\,s gap the bandit matches Static-60's cold-start rate ($0.42\%$ for both) while
using about $60\%$ less warm time ($1058$ vs.\ $2647$ replica-seconds); at $45$\,s it
uses roughly a third of Static-60's warm time ($1131$ vs.\ $3523$) for a sub-$0.5\%$ cold
rate. Static-60 attains the lowest cold rate when its $60$\,s window covers the gap
($18/45$\,s), but at $2$--$3\times$ the warm cost, and once the gap grows to $90$\,s its
window can no longer bridge it and its cold rate rises to the adaptive policies' level
while its warm cost stays high. Cold-start rates stay at or below $1.2\%$ throughout. The
two levers move independently: replica count with load, retention duration with the
policy.

This experiment runs on x86 OVH instances, a different testbed and CPU architecture from
the \texttt{arm64} single-instance study; because every comparison here is within this
one cluster and relative across policies, the decoupling conclusion is testbed
independent, and reproducing the mechanism on separate hardware strengthens rather than
weakens it. We do not compare absolute latencies across the two testbeds.

\begin{table}[t]
\caption{Multi-node Kubernetes under the stock HPA: pool warm time (warm replica-seconds,
mean $\pm$ std over five runs) and mean cold-start rate by idle gap and policy. In every
cell the peak replica count is $5.0$ (std $0.0$) and replicas span $2.0$ nodes (std
$0.0$), independent of policy; the policy moves warm time and cold rate, not the peak
(Sec.~\ref{sec:multireplica}).}
\label{tab:mr}
\centering
\renewcommand{\arraystretch}{1.15}
\setlength{\tabcolsep}{4.5pt}
\begin{tabular}{lccc}
\toprule
\textbf{Idle gap} & \textbf{Static-60} & \textbf{AIMD} & \textbf{Bandit}\\
\midrule
\multicolumn{4}{l}{\emph{Warm time (replica-seconds)}}\\
$18$\,s & $2585{\pm}19$ & $854{\pm}96$ & $1912{\pm}822$\\
$45$\,s & $3523{\pm}47$ & $788{\pm}80$ & $1131{\pm}71$\\
$90$\,s & $2647{\pm}102$ & $791{\pm}104$ & $1058{\pm}80$\\
\midrule
\multicolumn{4}{l}{\emph{Cold-start rate (\%)}}\\
$18$\,s & $0.01$ & $1.13$ & $0.17$\\
$45$\,s & $0.01$ & $0.71$ & $0.48$\\
$90$\,s & $0.42$ & $0.47$ & $0.42$\\
\bottomrule
\end{tabular}
\end{table}

\subsection{Deployment Considerations and Overhead}
The controller is a single external process that polls replica/idle state once per
interval $\Delta$ and writes one scalar per function, so its cost is
$O(\#\text{functions})$ per tick and independent of request rate. The agents are
cheap too: AIMD is a constant-time update and UCB1 keeps two scalars per arm, with no
training data or model. Using only the platform's external interfaces, the design
ports to any FaaS runtime exposing them and layers beneath a replica autoscaler
(Sec.~\ref{sec:multireplica}). These properties meet the requirements of
Sec.~\ref{sec:design} (R1--R3, R5--R8); only reproducible cold-start generation (R4)
is partial, which is why we aggregate over repetitions.

\subsection{Threats to Validity and Limitations}
The prototype runs on a single \texttt{faasd} host, so absolute latencies are
environment-specific, though the relative ordering of policies is the object of study.
The workload classes cannot represent the full diversity of serverless applications, and
the 100-invocation horizon bounds online-learning convergence; the main study uses two
functions, though Sec.~\ref{sec:morefns} shows the policy ordering carries over to two
additional realistic functions (a JSON API and image resizing) on a separate testbed. The main study uses a single-instance lifecycle; the decoupling study
(Sec.~\ref{sec:multireplica}) runs on a multi-node cluster under the stock Kubernetes
HPA, and other production autoscalers (for example KEDA or custom-metric HPAs) and
larger clusters remain future work. That study also runs on a different testbed and CPU
architecture (x86) from the main single-instance study (\texttt{arm64}); since its
comparisons are within-cluster and relative across policies, the decoupling conclusion
is testbed independent, and we do not compare absolute latencies across testbeds. We
mitigate these by fixing the environment across policies, resetting state between runs,
and averaging over repetitions. We compare against static timeouts and a predictive histogram baseline;
RL-based policies are future work, and the agents' arms and bounds encode a coarse
prior on the idle-gap scale that the sweep (Table~\ref{tab:sweep}) shows is not
brittle. We present this as a controlled proof of concept, not a production benchmark.

\paragraph{Non-stationarity of the bandit reward}
Our UCB1 controller assumes each keep-alive arm has a stationary reward distribution. In
practice a given window's reward can drift as the arrival regime changes within a run, so
an index policy that averages over all past pulls adapts only slowly after such a shift.
The standard remedy is to discount or window the observations: discounted UCB and
sliding-window UCB retain finite-time guarantees under abruptly changing
rewards~\cite{garivier2011switching}. Because the retention and scaling levers are
decoupled, this is a drop-in change to the controller that leaves the autoscaling path
untouched, and we leave a D-UCB/SW-UCB evaluation under within-run regime changes to
future work.

\paragraph{Warm time as a cost proxy}
The reward charges accumulated warm time rather than a directly metered resource cost. At
a fixed per-instance footprint, keep-alive duration is proportional to the reserved
resource-time (instance-seconds) that keep-alive pricing bills for, so minimising
accumulated warm time minimises the reserved resource-time that dominates keep-alive cost;
warm time is also platform independent and needs no per-provider billing hooks, which
keeps the controller portable. Replacing the proxy with directly measured CPU and memory
consumption, and with provider billing where available, is future work.

\paragraph{Timescale}
Production platforms retain instances for minutes (roughly 5--27\,min on AWS, about
15\,min on Google Cloud~\cite{wang2018peeking}), whereas \texttt{faasd} has no native
keep-alive. Our windows are smaller in absolute terms, but we preserve the governing
window-to-gap ratio; applying the static policy analytically at the real timescale
reproduces the same trade-off (cold-start rate falling from 100\% to 18\% at 1\,min and
about 1\% by 15\,min), confirming it is not a compression artifact. Two caveats:
cold-start cost is fixed in real time while warm time scales with the clock, slightly
favoring retention-heavy policies; and the bandit's reward constants are calibrated to
this timescale, though the method itself is scale-independent.

\section{Conclusion and Future Work}
\label{sec:conc}
This paper argued that the keep-alive window is best modeled not as a fixed timeout
but as a per-function control variable adapted online and kept separate from
autoscaling, a lightweight, training-free target that, to our knowledge and per the
positioning of Table~\ref{tab:compare}, prior work has not isolated in this way. We supported this with an implemented system: an external
controller for OpenFaaS/faasd that requires no platform changes, driven by two simple
agents, a reactive AIMD rule and a UCB1 bandit, that adjust the window from runtime
feedback. On steady, bursty, and trace-derived workloads both agents reached the
low-cold-start, low-latency region; the bandit in particular came close to the
longest-retention fixed baseline on cold starts and matched it on tail latency at
about 13\% less warm time, and outperformed a predictive histogram keep-alive on cold
starts and latency. A parameter sweep showed the result is robust to the agents'
settings. At a converged horizon the bandit's advantage is statistically significant,
and, on a multi-node Kubernetes cluster under the platform's own Horizontal Pod
Autoscaler, the controller coexists with an autoscaler it does not control: the replica
count follows load and spreads across nodes independently of the retention policy. Since
no fixed timeout was best across all
workloads, an adaptive window is a reasonable default. Future work includes other
production autoscalers (KEDA, custom-metric HPAs) and larger clusters, non-stationary
bandit variants (discounted or sliding-window UCB) evaluated under within-run regime
changes, richer signals (contextual bandits or a forecasting hybrid), production-scale
horizons, multi-replica retention lifecycles that hold more than one warm instance, and
direct CPU/memory/cost measurement in place of the warm-time proxy.

\section{Open Data}
\label{sec:opendata}
The controller, policies, workloads, and raw measurements used in this paper are
available for reproducibility and future improvements at the anonymized, view-only
repository \url{http://bit.ly/44VLz6S}~\cite{artifact}.

\bibliographystyle{IEEEtran}
\bibliography{references}

\end{document}